\documentclass[%
 reprint,
 onecolumn,
 amsmath,amssymb,
 aps,
11pt,
]{revtex4-2}

\usepackage{graphicx}% Include figure files
\usepackage{dcolumn}% Align table columns on decimal point
\usepackage{bm}% bold math
\usepackage[colorlinks,linkcolor=blue,anchorcolor=blue,citecolor=blue,urlcolor=black]{hyperref}
\usepackage[mathlines]{lineno}% Enable numbering of text and display math
\begin{document}

\preprint{APS/123-QED}

\title{Pattern preservation during the decay and growth of localized wave packet \\ in two-dimensional channel flow}% Force line breaks with \\

\author{Lin-Sen Zhang}
\author{Jian-Jun Tao}%
 \email[Corresponding author: ]{jjtao@pku.edu.cn}
\affiliation{%
HEDPS-CAPT, SKLTCS, Department of Mechanics and Engineering Science, College of Engineering, Peking University,
Beijing, 100871, P. R. China
}%
 %\author{co-author}

%\affiliation{
%\affiliation{
%\affiliation{%

\date{\today}% It is always \today, today,

\begin{abstract}
In this paper, the decay and growth of localized wave packet (LWP) in two-dimensional plane-Poiseuille flow are studied numerically and theoretically. When the Reynolds number ($Re$) is less than a critical value $Re_c$,  the disturbance kinetic energy $E_k$ of LWP decreases monotonically with time and experiences three decay periods, i.e. the initial and the final steep descent periods, and the middle plateau period. Higher initial $E_k$ of a decaying LWP corresponds to longer lifetime. According to the simulations, the lifetime scales as $(Re_c-Re)^{-1/2}$,  indicating a divergence of lifetime as $Re$ approaches $Re_c$, a phenomenon known as ``critical slowing-down''. By proposing a pattern preservation approximation, i.e. the integral kinematic properties (e.g. the disturbance enstrophy) of an evolving LWP are independent of $Re$ and single valued functions of  $E_k$,  the disturbance kinetic energy equation can be transformed into the classical differential equation for saddle-node bifurcation, by which the lifetimes of decaying LWPs can be derived, supporting the $-1/2$ scaling law. Furthermore, by applying the pattern preservation approximation and the integral kinematic properties obtained as $Re<Re_c$, the Reynolds number and the corresponding $E_k$ of the whole lower branch, the turning point, and the upper-branch LWPs with $E_k<0.15$ are predicted successfully with the disturbance kinetic energy equation, indicating that the pattern preservation is an intrinsic feature of this localized transitional structure. 
%\begin{description}
%\end{description}
\end{abstract}

\maketitle

%\tableofcontents

\section{Introduction}\label{sec:introduction}
Localized structures are found to be the key characteristics of subcritical transitions in wall-bounded shear flows \cite{Coles65, Atta66, tuckerman2020patterns}, e.g. puffs in pipe flows and turbulent bands in plane Poiseuille flow (PPF), plane Couette flow (PCF), and circular Couette flow. For the two-dimensional PPF, the subcritical transition may occur at Reynolds numbers ($Re$) larger than 89, the threshold for energy instability \cite{Orr1907, Schmid07}, and less than 5772, the critical value of linear instability \cite{Orszag71}, and the counterpart of turbulent band is localized Tollmien-Schlichting waves or localized wave packet (LWP) \cite{jimenez1990transition, price1993numerical, Teramura16, Barnett17}. A possible localization mechanism is the subharmonic instability \cite{drissi1999subharmonic, mellibovsky2015mechanism} of upper-branch Tollmien-Schlichting waves (TSW) \cite{Chen73, Zahn74}, which do not exist for Reynolds numbers less than 2600 \cite{Soibelman91, drissi1999subharmonic, Ayats20}. By applying the continuation of invariant solution in a periodic domain with a periodicity length $L$ about 16.7$\pi$, the modulated Tollmien-Schlichting waves (MTSWs) generated through Hopf bifurcation from TSW is tracked to a critical Reynolds number $Re_c$ about 2285 \cite{mellibovsky2015mechanism}. When the periodicity length is increased to $32\pi$ , the minimal Reynolds number traced by the continuation method for LWP is found to be 2335 \cite{Zammert17}, above which the lower branch is unstable, corresponding to the edge state \cite{Skufca06}, and the upper branch undergoes a Hopf bifurcation at a higher Reynolds number. With time-dependent numerical simulations in a long channel ($L=20\pi$), single localized wave packet was obtained with  $Re_c \sim 2330$ \cite{price1993numerical}, indicating a saddle-node bifurcation. Such kind of bifurcation is also found in PCF \cite{nagata1990three} and pipe flow \cite{Faisst03}. Later simulations of two-dimensional PPF reveal $Re_c$ of 2332 with finer resolutions and much longer channel lengths, e.g. $L=1600$ \cite{wang2015study} and $L=400$ \cite{Xiao2021self}, indicating that the periodic MTSW obtained previously as $Re>2332$ will evolve to LWP when the channel is long enough.

An important issue for subcritical transition is the persistency of localized structures, which can be determined by analyzing their lifetimes $\tau$. For PCF, it is shown experimentally that the lifetime of turbulent patches scales as $\tau \propto (Re_c-Re)^{-1}$ with $Re_c \approx 323 \pm 2$ \cite{bottin1998statistical}, which agrees with the numerical observation $Re_c = 324\pm 1$ \cite{Duguet10}. The same scaling relation was found for puffs in the pipe flows by experiments \cite{peixinho2006decay} and numerical simulations \cite{willis2007critical}, representing that the localized turbulence can sustain when $Re>Re_c$. Later experiments with longer observation time showed that the mean lifetimes of puff before decaying and splitting are superexponential functions of $Re$, indicating that puffs are transient phenomena at moderate Reynolds numbers \cite{hof2008repeller, avila2011the}. The lifetime statistics of turbulent bands in tilt and narrow computational domains have been studied as well, and the Reynolds numbers determined by the crossover of the mean decay time and splitting time for PPF and PCF are 965 \cite{gome2020statistical} and 325 \cite{shi2013scale}, respectively. For large channel domains, however, it is illustrated numerically that the transition starts with a sparse turbulence state, where the localized turbulent bands are sustained by a dynamic balance between band extension and band breaking \cite{xiong2015turbulent, tao2018extended}.  The lower bound or threshold Reynolds number of the sparse turbulence state is $Re_c=660$, which is confirmed experimentally by velocity field measurements  \cite{Liu20entropy, liu2020extension} and flow visualizations \cite{Mukund21}. Above $Re \approx 700$, most turbulent bands are found numerically to have the same orientation and the spanwise symmetry is restored and the two-dimensional directed-percolation is retrieved only above $Re\approx 1000$ \cite{shimizu2019bifurcations}. For two-dimensional PPF, however, the scaling law of LWP's lifetime with $Re$ hasn't been studied so far.

The spatio-temporal evolution of localized flow structures, e.g. decay and growth, is another important feature of the subcritical transitions. For PCF, the decay of  turbulent band was studied in large domains with under-resolution simulations, and it was found that the turbulent band shrinks slowly and laminar holes are formed in the bands \cite{Manneville11, Rolland15}.  At moderate Reynolds numbers, direct numerical simulations illustrate that an isolated turbulent band decays in a style of longitudinal contraction with a statistically constant velocity, and experiences transient growths before the eventual relaminarization \cite{Lu19}. In PPF, turbulent band breaking occurs and the fragments relax to streaks before vanishing \cite{xiong2015turbulent}. The growth of transitional characteristic structures has a plenty of scenarios. An isolated turbulent band in PCF may experience oblique extension, transverse split, and longitudinal split at moderate Reynolds numbers \cite{Lu21}, and their combined effects lead to a labyrinthine growth \cite{Manneville12}, which can also be triggered directly by localized perturbations \cite{Duguet10}. For the turbulent bands in PPF, the oblique extension happens at moderate Reynolds numbers, and the parallel or longitudinal split and the transverse split occur at relatively high Reynolds numbers \cite{xiong2015turbulent, shimizu2019bifurcations}. In the two-dimensional PPF, however, it is still unknown whether LWPs evolve in similar ways to the three-dimensional cases during the decay and growth processes.

This paper is organized as follows. In Sec. \ref{sec: numerical method} the numerical methods are described, and the resolution and domain size independence are verified for LWP. Different decaying periods are revealed at low Reynolds in Sec. \ref{sec: decay}, and the lifetime scaling law is explained theoretically with a pattern preservation approximation and the disturbance kinetic energy equation in Sec. \ref{sec: pattern preservation}. The growth of LWP from the lower branch or edge state  to the upper branch is studied numerically and compared with theoretical predictions in Sec. \ref{sec: growth}, and some concluding remarks are provided in Sec. \ref{sec: conclusions}.

\section{Physical model and numerical methods} \label{sec: numerical method}
The incompressible two-dimensional plane Poiseuille flow is numerically simulated with a spectral code \cite{SIMSON}. The flow rate is kept constant, and the half channel height $h$ and 1.5 times of the bulk velocity $U_m$ are chosen as the characteristic length and velocity scales, respectively. No-slip boundary conditions at the side walls ($y=\pm 1$) and periodic boundary condition in the streamwise direction $x$ are applied. $N_x$ Fourier modes and $N_y$ Chebyshev modes are used in the $x$ and $y$ directions, respectively. The Reynolds number is defined as $Re=1.5U_mh/\nu $, where $\nu$ is the kinematic viscosity of the fluid. For details of the simulation methods, we refer to the previous papers \cite{xiong2015turbulent, tao2018extended}.

Following the previous studies \cite{Xiao2021self}, the length $l_p$, the center coordinate $x_c$, and the convection velocity $c_p$ of LWP are defined as

\begin{equation}\label{eq:lp}
l_p = 2\sqrt{3}\left[\frac{\int e_kx^2 dxdy}{\int e_kdxdy} - \left(\frac{\int e_kx dxdy}{\int e_kdxdy}\right)^2\right]^{1/2},\ x_c = \frac{\int e_k x dxdy}{\int e_k dxdy},\ c_p = \frac{dx_c}{dt},
\end{equation}
where $e_k$ is the kinetic energy of disturbance velocity ${\bf{u'}}=(u',v')=(u-U_0,v')$ and $U_0(y)=1-y^2$ is the basic flow solution.

%\section{Saturated LWP}\label{sec:real}

\begin{figure*}
	\centerline{\includegraphics[scale=0.125]{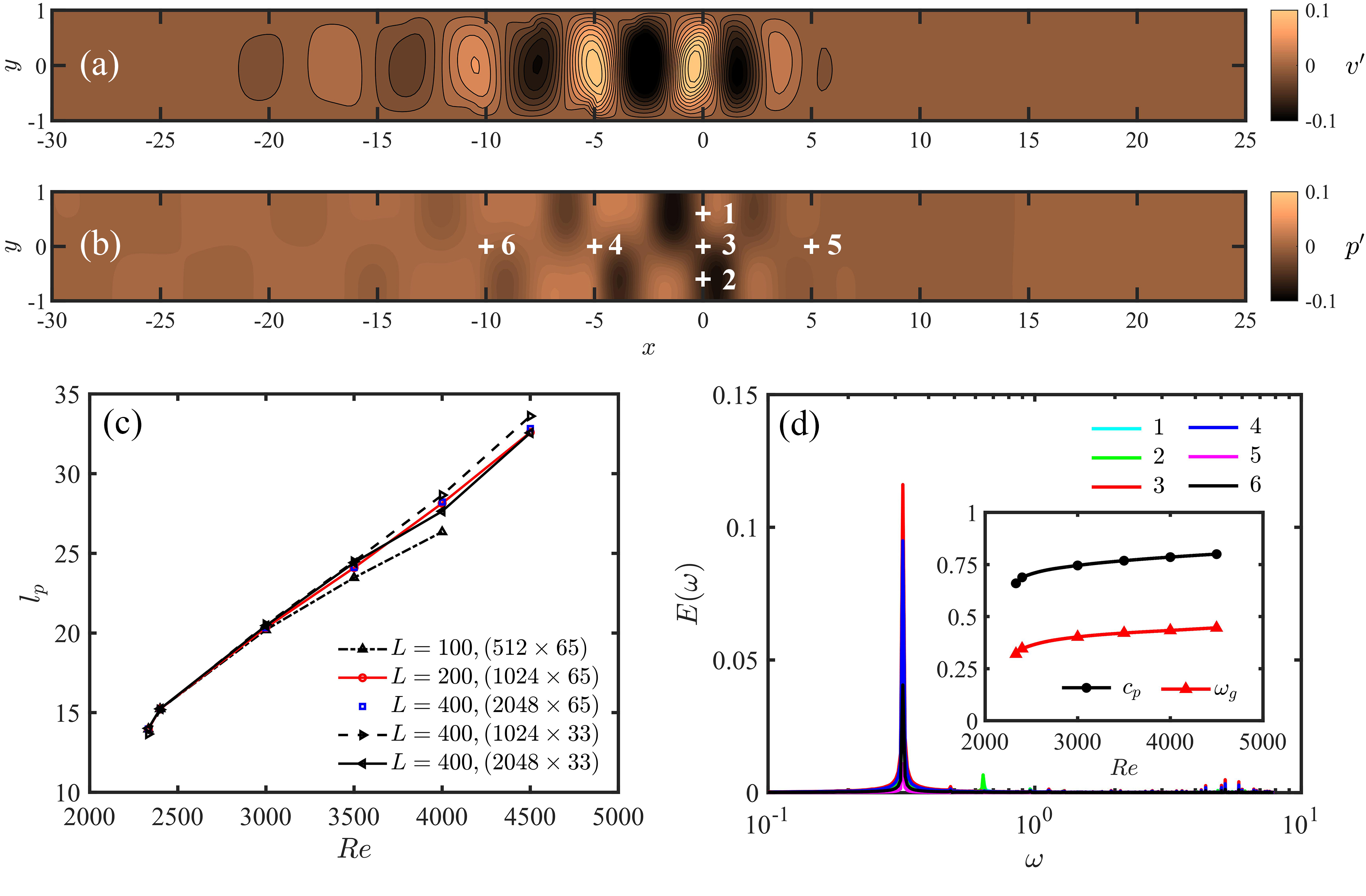}}
	\caption{(a) Transient normal velocity field and (b) disturbance pressure field of a LWP obtained at $Re=2333$. (c) Length of LWP $l_p$ obtained at different $Re$s with different domain lengths $L$ and numerical resolutions $(N_x\times N_y)$. (d) Power spectra of $v'$ measured in a frame moving with LWP's convection velocity $c_p=0.66$. The measurement locations numbered by 1 to 6 are shown in (b). The global angular frequency $\omega_g$ and $c_p$ as functions of $Re$ are shown in the inset of (d).}
	\label{fig1}
\end{figure*}

The computational domain length $L$ is just the space interval between neighbor LWPs due to the periodic boundary condition, and its effects on $l_p$ are shown in Fig. \ref{fig1}, where the same initial field shown in Fig. \ref{fig1}(a) is used. With the increase of $L$ and the spatial resolution ($N_x\times N_y$), $l_p$ converges at different Reynolds numbers [see Fig. \ref{fig1}(c)], reflecting the independence of LWP on the computational domain size and resolution. In the previous studies, ($N_x\times N_y$)=$(200\times33)$ with $L=20\pi$ \cite{price1993numerical} and $(256\times33)$ per 100 units length \cite{wang2015study} were used. According to Fig. \ref{fig1}, larger $L$ and higher spatial resolution are required to obtain converged solutions at higher $Re$s. For example, $l_p$ is about 33 at $Re=4500$, and $L=100$ is so short that the interaction among the LWP and its neighbors leads to decay, a similar phenomenon is observed for turbulent bands in three dimensional channel flows \cite{tao2018extended}. Consequently, $L=400$ and $(N_x\times N_y) = (2048\times65)$ are used hereafter. In a translational coordinate moving with the LWP's velocity $c_p$, the time series of normal velocities $v'$ are recorded at different locations  shown by the symbols in Fig. \ref{fig1}(b) , and the corresponding power spectra are calculated. These spectra share the same dominant frequency [see Fig. \ref{fig1}(d)], which is identified as a global frequency $\omega_g$ by spatio-temporal stability analyses \cite{Xiao2021self} and increases mildly with $Re$ in a similar manner to $c_p$ as shown in the inset. 

\section{Decay of LWP}\label{sec: decay}

\begin{figure*}
	\centerline{\includegraphics[scale=0.125]{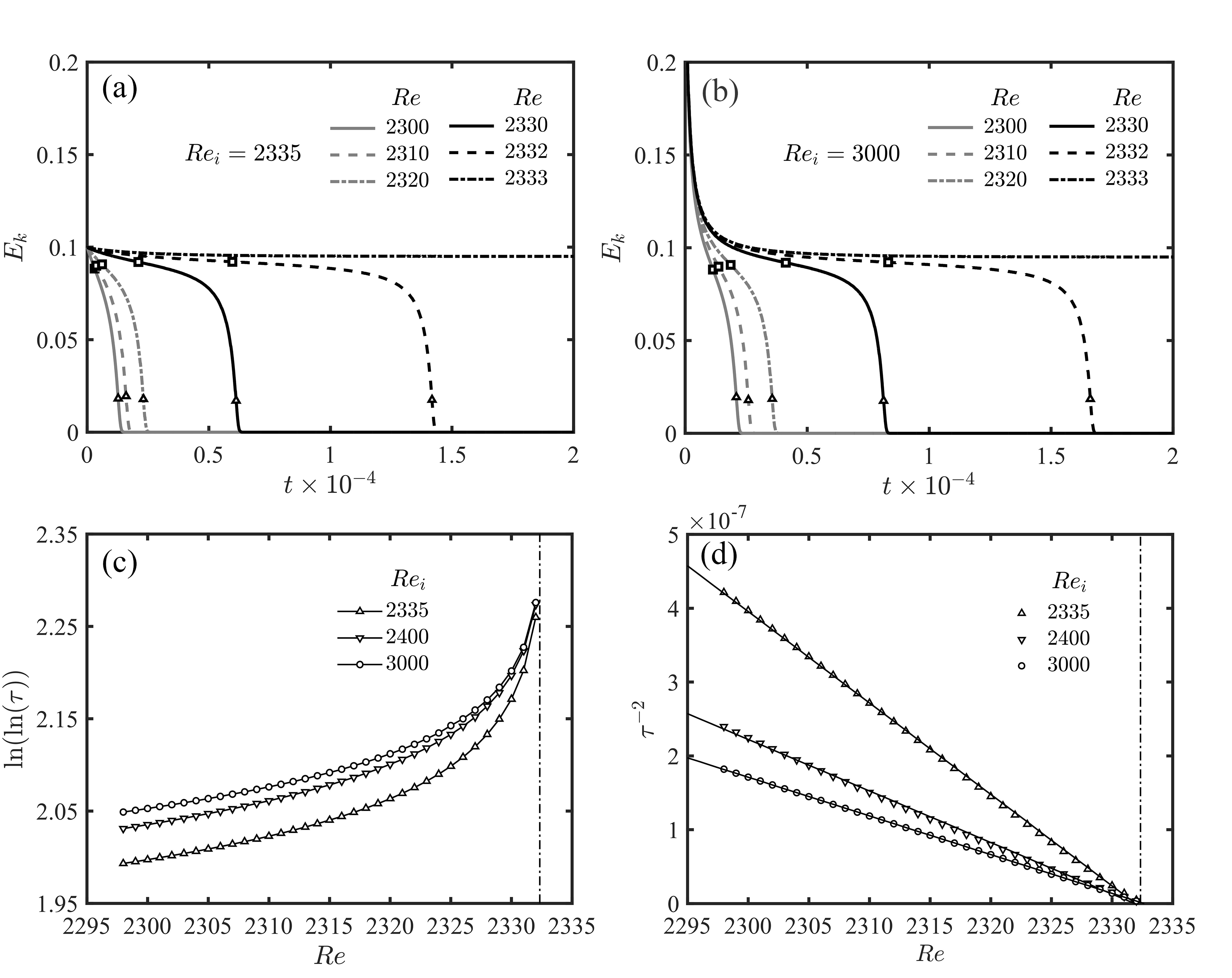}}
	\caption{The temporal evolutions of disturbance kinetic energy $E_k$ at different Reynolds numbers are shown in (a) and (b) with different initial disturbances, the LWPs obtained at $Re_i$s. The symbols $\square$ and $\triangle$ indicate the first and second inflection points, respectively. (c) The lifetimes $\tau $ of different initial LWPs ($Re_i$s) as functions of $Re$. (d) $\tau^{-2}$ as a function of $Re$. The vertical dash-dot lines in (c) and (d) represent $Re =2332.35$.}
	\label{fig2}
\end{figure*}

In order to study the decay properties and lifetime characteristics of LWP, the LWPs obtained at $Re =$ 2335, 2400, and $3000$ are used as the initial disturbances of simulations within the range of $Re = 2298\sim 2333$. The total disturbance kinetic energy $E_k = \int e_k dxdy$ is calculated, and the lifetime $\tau$ of each case is defined as the period with $E_k\ge  10^{-5}$. Though the initial disturbance kinetic energies are different, it is shown in Fig. \ref{fig2} that the temporal evolutions of $E_k$ illustrate some common features. Firstly,  $E_k$ obtained at $Re=2333$ reaches a constant value eventually, while other cases at lower $Re$s decay monotonically: the higher $Re$ is, the longer the LWP lifetime is. Secondly, for LWPs with high initial $E_k$, e.g. the cases shown in Fig. \ref{fig2}(b), there are three characteristic periods along the $E_k$ evolution curves, i.e. the initial steep descent period, the middle plateau period, and the final steep descent period. The last two periods are characterized by the first and the second inflection points  labeled by the square and the triangle symbols in Figs. \ref{fig2}(a) and \ref{fig2}(b), respectively. $E_{k1}$ and $E_{k2}$ are the corresponding disturbance kinetic energies. Finally, the lifetime increases faster than the superexponential scaling when $Re$ is close to a critical value $Re_c$ indicated by a vertical dash dot line in Fig. \ref{fig2}(c). Note that the lifetimes of transient puffs in pipe flows scale superexponentially with $Re$ \cite{hof2008repeller}. As shown in Fig. \ref{fig2}(d), $\tau^{-2}$ of different initial disturbances decreases linearly and can be extrapolated to a value $Re_c = 2332.35\pm 0.35$, i.e. $\tau^{-2} \propto (Re_c-Re)$ or $\tau \propto (Re_c-Re)^{-1/2}$, indicating that the lifetime tends to infinity when $Re$ increases to $Re_c$ and LWP becomes a temporally persistent structure as $Re>Re_c$. This critical Reynolds number is consistent with the previous values obtained with initial random disturbances \cite{wang2015study} and localized disturbances \cite{Xiao2021self}, where LWPs are arranged initially with large intervals.

\section{Pattern preservation and decay mechanisms } \label{sec: pattern preservation}

\begin{figure*}
	\centerline{\includegraphics[scale=0.125]{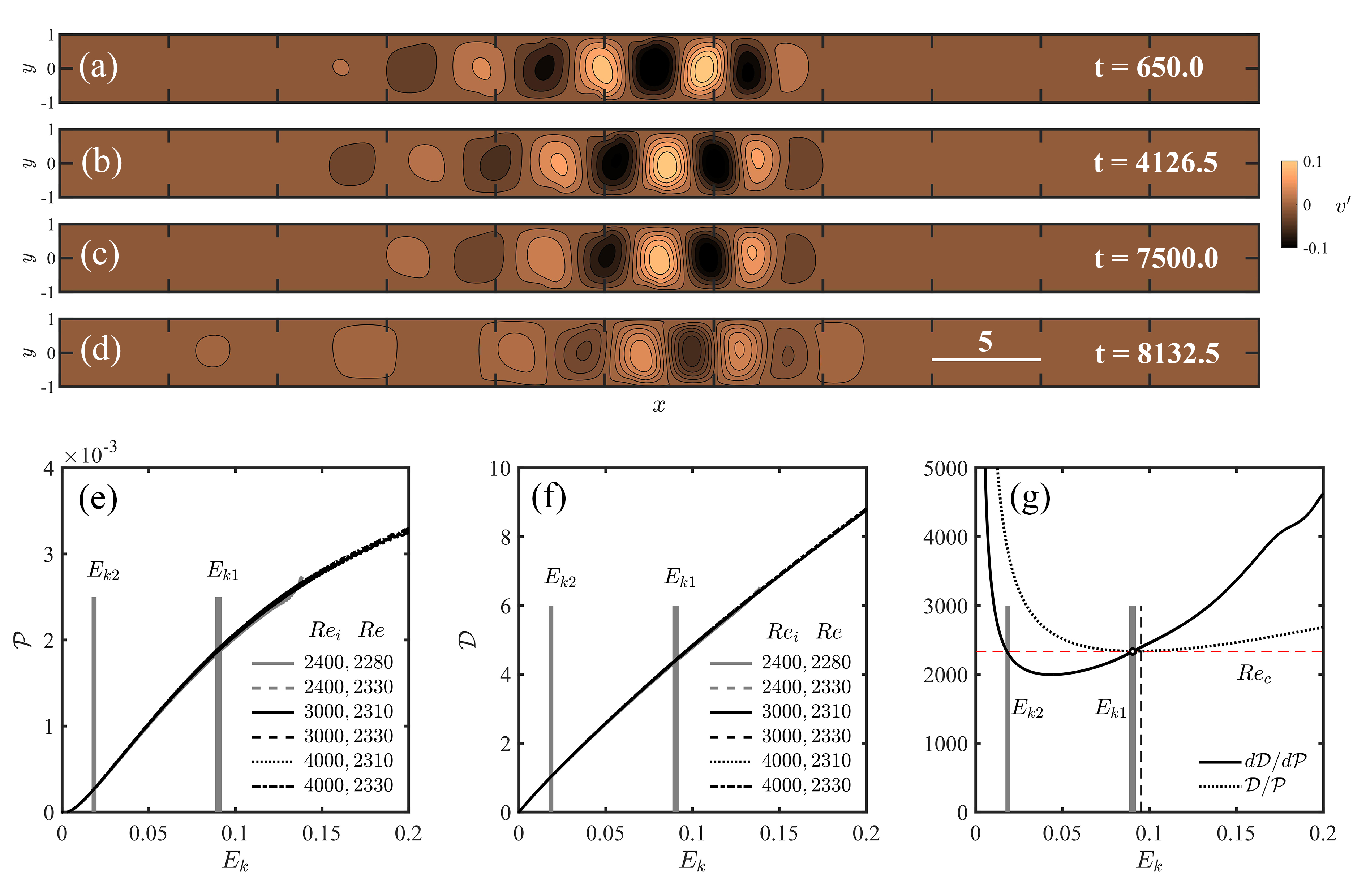}}
	\caption{The iso-contours of $v'$ obtained at different times (a-d) for $Re=2330$. The initial disturbance is the LWP obtained at $Re_i =3000$. The energy production term $\mathcal{P}$ and the disturbance enstrophy $\mathcal{D}$ in Eq. (\ref{eq51}) calculated for different $Re$s with different initial LWPs (obtained at $Re_i$) are shown in (e) and (f), respectively. (g) $\mathcal{D}/\mathcal{P}$ and $d \mathcal{D}/d \mathcal{P}$ as functions of $E_k$ for the decaying case shown in (a-d). The vertical dashed line and the circle represent $E_{kc}=0.095$ and the point satisfying $d\mathcal{D}/d\mathcal{P}=Re_c$, respectively. The disturbance kinetic energies at the inflection points, $E_{k1}$ and $E_{k2}$ obtained at different $Re$s with different initial disturbances, are shown as vertical grey bars in (e-g). }
	\label{fig3}
\end{figure*}

The lifetimes of LWPs change for different Reynolds numbers and initial disturbances, but the common features shown in Fig. \ref{fig2} suggest that the decay process is completely different from the chaotic behaviors of puffs in pipe flows. It is shown in Figs. \ref{fig3}(a-d) that the flow patterns obtained at different decay moments look similar, i.e. localized wave packets whose streamwise wavelengths do not change much with time but whose perturbation amplitudes and kinetic energy decrease monotonically. In addition, $E_k$ at the inflection points on the $E_k$ curves are nearly the same for different Reynolds numbers and different initial perturbation energies [see Figs. \ref{fig2}(a) and \ref{fig2}(b)]. Therefore, LWPs seem to follow the same route to decay, preserving the flow pattern but fading the kinetic energy, and hence a pattern preservation approximation is proposed: for an evolving LWP, the domain integral properties of the velocity field do not depend on the Reynolds number and the initial perturbation strength, but are single valued functions of $E_k$.

Applying the no-slip boundary conditions on the sidewalls and periodic conditions in the streamwise direction, the growth rate $\sigma$ of $E_k$ is defined according to the Reynolds-Orr equation,
\begin{equation}\label{eq51}
\sigma=\frac{dE_k}{dt} = \mathcal{P}-\frac{1}{Re}\mathcal{D},\  \mathcal{P}=-\int u'v'\frac{dU_0}{dy} dxdy,\  \mathcal{D}= \int\nabla \mathbf{u'}:\nabla \mathbf{u'} dxdy=\int (\nabla \times \mathbf{u'})^2 dxdy,
\end{equation}
where $\mathcal{P}$ and $\mathcal{D}$ are the kinetic energy production term and the enstrophy of the disturbance field, respectively, and $\mathcal{D}/Re$ represents the viscous dissipation.

Along the decay processes, both $\mathcal{P}$ and $\mathcal{D}$ are calculated for different Reynolds numbers and different initial kinetic energies and are shown in Figs. \ref{fig3}(e) and \ref{fig3}(f), respectively. The curves collapse with each other, indicating that both $\mathcal{P}$ and $\mathcal{D}$ are single-valued, monotonically increasing functions of $E_k$. Consequently, the pattern preservation approximation is validated. It is noted that the $\mathcal{D}$ and $\mathcal{P}$ curves shown in Fig. \ref{fig3} have small-amplitude oscillations, which are related to the periodic production of travelling waves at the downstream end of LWP \cite{Xiao2021self} during the decay process. In order to eliminate the interference of these oscillations, the growth rate $\sigma$, $\mathcal{D}/\mathcal{P}$, and $d\mathcal{D}/d\mathcal{P}$ are calculated with smoothed $E_k$ curves hereafter, where the oscillation components have been filtered out.

When $Re$ is close to $Re_c$, or $Re=Re_c(1-\delta )$ with $0<\delta \ll 1$, we have
\begin{equation}\label{eq52}
\sigma = \mathcal{P} - \frac{1}{Re}\mathcal{D}=\mathcal{P} - \frac{1}{Re_c(1-\delta)}\mathcal{D}=\mathcal{P} - \frac{1}{Re_c}\mathcal{D}[1+\delta+O({\delta}^2)].
\end{equation}
If a LWP is saturated at $Re_{sat}$, we have $\sigma=0$ and $\mathcal{D}/\mathcal{P}=Re_{sat}$ according to Eq. (\ref{eq51}). Considering that $\sigma=0$ at $Re_c$ and  $Re_c$ is the minimum of $Re_{sat}$, where $dRe_{sat}/dE_k = d(\mathcal{D}/\mathcal{P})/dE_k=0$, or $d\mathcal{D}/d\mathcal{P} = \mathcal{D}/\mathcal{P}$, the subcritical transition threshold $Re_c$ can be derived analytically based on the $Re$-independent properties ($\mathcal{D}$ and $\mathcal{P}$) as
\begin{equation}\label{eq53add}
Re_c = \frac{\mathcal{D}}{\mathcal{P}} \text{ where } \frac{d\mathcal{D}}{d\mathcal{P}} = \frac{\mathcal{D}}{\mathcal{P}}.
\end{equation}
As shown in Fig. \ref{fig3}(g), the crossover point of $d\mathcal{D}/d\mathcal{P}$ (thick solid line) and $\mathcal{D}/\mathcal{P}$ (thick dotted line) is consistent with $Re_c$ obtained in simulations [see Fig. \ref{fig2}(d)]. 
\begin{figure*}
	\centerline{\includegraphics[scale=0.125]{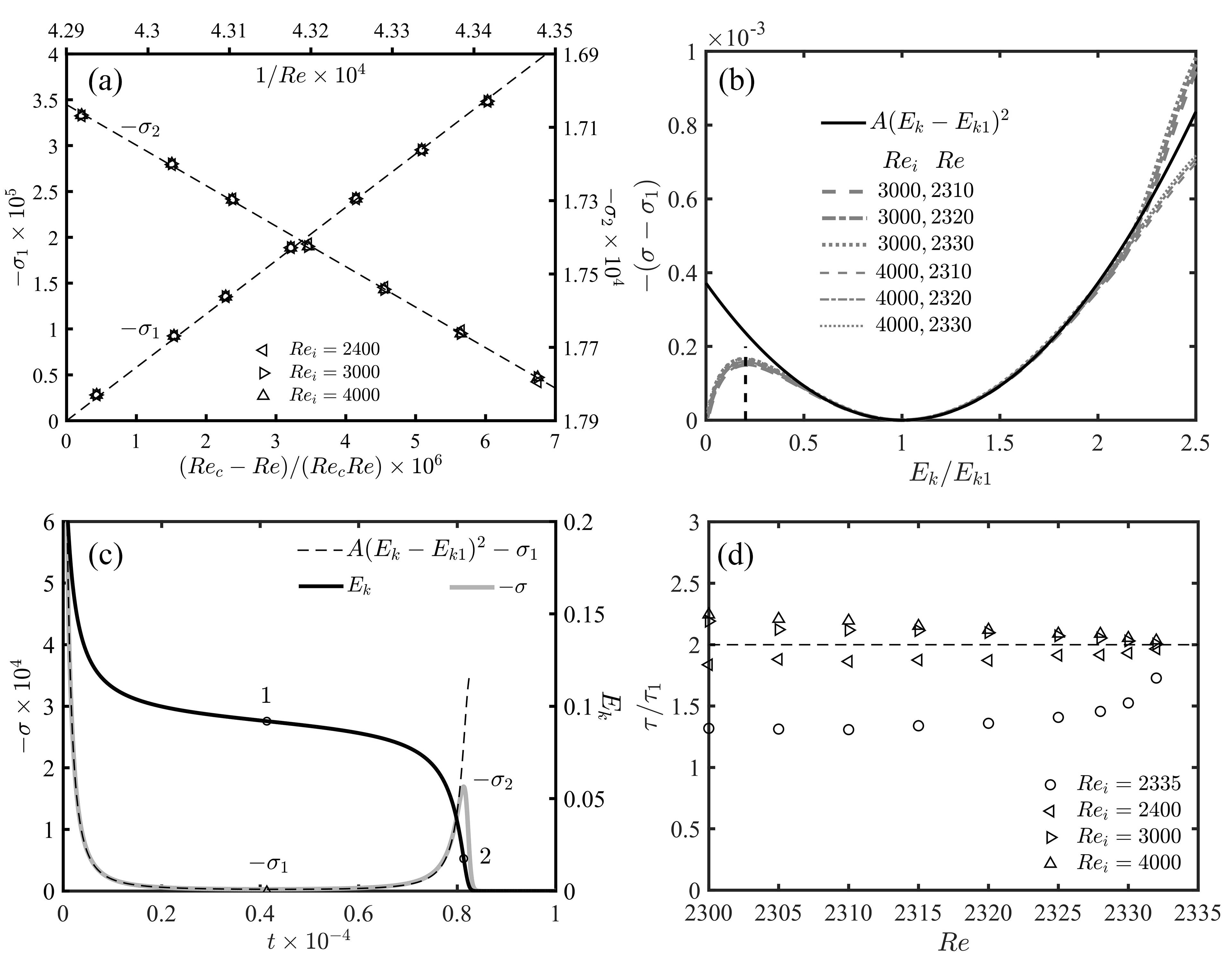}}
	\caption{Decay properties obtained at different $Re$s with different initial LWPs obtained at $Re_i$. (a) The decay rates at the inflection points, $-\sigma_1$ and $-\sigma_2$. (b) Decay rate difference $-(\sigma-\sigma_1)$ as a function of $E_k/E_{k1}$. $A=0.044$, agreeing well with $-(1/2)d^2\sigma/dE_k^2(E_{kc})=0.048$ as $1/Re \simeq 1/Re_c$ is used. The vertical dashed line indicates $E_k=E_{k2}$. (c) $-\sigma$ and $E_k$ as functions of $t$ at $(Re_i, Re)=(3000, 2330)$. (d) Lifetime ratio $\tau/\tau_1$ as a function of $Re$, where $\tau_1$ is the period as $E_k\le E_{k1}$.
}
	\label{fig4}
\end{figure*}

When $Re < Re_c$, $dE_k/dt < 0$ and the inflection points in the $E_k$ curves require
\begin{equation}\label{eq53}
 \frac{d^2 E_k}{dt^2} = \left(\frac{d\ \mathcal{P}}{dE_k} -\frac{1}{Re}\frac{d\ \mathcal{D}}{dE_k}\right)\frac{dE_k}{dt} =0, \text{ or  } \frac{d\ \mathcal{D}}{d\ \mathcal{P}}  = Re = Re_c-\delta Re_c \simeq Re_c.
\end{equation}
Considering that $d\mathcal{D}/d\mathcal{P}$ only depends on $E_k$, Eq. (\ref{eq53}) suggests that the $E_k$s at the inflection points ($E_{k1}$ and $E_{k2}$) are nearly constants, a feature confirmed by the numerical data shown in Figs. \ref{fig2}(a) and \ref{fig2}(b). As illustrated by the vertical dashed line in Fig. \ref{fig3}(g),  $E_{k1}\simeq E_{kc}$, a result consistent with Eqs. (\ref{eq53add}) and (\ref{eq53}). Consequently, the growth rate at the first inflection point is
\begin{equation}\label{eq54}
\sigma_1 \simeq \sigma(E_{kc})  = \mathcal{P}(E_{kc}) - \frac{1}{Re}\mathcal{D}(E_{kc})=\frac{1}{Re_c}\mathcal{D}(E_{kc}) - \frac{1}{Re}\mathcal{D}(E_{kc})\propto \frac{Re_c-Re}{Re_cRe},
\end{equation}
a relation confirmed by simulations shown in Fig. \ref{fig4}(a).

 Considering the growth rate at the second inflection point  $\sigma_2 = \mathcal{P}(E_{k2}) - \mathcal{D}(E_{k2})/Re$, a linear relation between $\sigma_2$ and $1/Re$ is expected and is confirmed by the numerical data shown in Fig. \ref{fig4}(a), where the fitted dashed line represents $\sigma_2=0.000382-1.288/Re$ and the coefficients agree reasonably with the values of $\mathcal{P}(E_{k2})$ and $-\mathcal{D}(E_{k2})$, 0.000304 and $-1.02$, respectively.

 When $Re$ approaches to $Re_c$, $\sigma$ in Eq. (\ref{eq52}) is approximately a single valued function of $E_k$  and can be expanded around the first inflection point as
\begin{equation}\label{eq55}
  \begin{split}
\sigma  &= \sigma_1 + \frac{d \sigma}{d E_k}\mid_{E_k=E_{k1}}(E_k-E_{k1}) + \frac{1}{2}\frac{d^2 \sigma}{d E_k^2}\mid_{E_k=E_{k1}}(E_k-E_{k1})^2 + ...\\
 &\simeq \sigma_1 + \frac{1}{2} \frac{d^2 \sigma}{d E_k^2}\mid_{E_k=E_{k1}}(E_k-E_{k1})^2 \\
 &\simeq \sigma_1 -A(E_k-E_{k1})^2,
   \end{split}
\end{equation}
where $A$ is approximated as a positive constant when $1/Re \simeq 1/Re_c$ is used. Note that at the inflection point, $(d \sigma /d E_k)\mid_{E_k=E_{k1}}=[(d \sigma /dt)/(d E_k/dt)]\mid_{E_k=E_{k1}}=0$.  When $\delta \rightarrow 0$, substituting Eq. (\ref{eq54}) into Eq. (\ref{eq55}) leads to a simplified equation for the reduced kinetic energy $E_k-E_{k1}$,
\begin{equation}\label{eqsdb}
\frac{d(E_k-E_{k1})}{dt} = \mathcal{D}(E_{kc})\frac{Re-Re_c}{Re_c^2} -A(E_k-E_{k1})^2,
\end{equation}
the classical partial differential equation for saddle-node bifurcation, i.e. when $Re<Re_c$ there are only decaying solutions for $E_k-E_{k1}$, and when $Re>Re_c$ there is a stable upper branch solution  and an unstable lower-branch one. After rescaling the time and the reduced kinetic energy with LWP's lifetime scale $\hat{\tau}$ as $t=\hat{\tau}\tilde{t}$ and $E_k -E_{k1} =\tilde{E_k}/(A\hat{\tau} )$, Eq. (\ref{eqsdb}) can be transformed into the unified form $d\tilde{E}_k /d\tilde{t}=-1-\tilde{E}_k^2$ for the decay cases by setting $1/\hat{\tau}^2=(Re_c-Re)A \mathcal{D}(E_{kc})/Re_c^2$, i.e. $\hat{\tau} \propto (Re_c-Re)^{-1/2}$, the lifetime scaling law found in the simulations shown in Fig. \ref{fig2}(d). Equation (\ref{eqsdb}) suggests that there will be long-lasting transient process (the middle plateau period) and the lifetime will diverge when the system is close to the saddle-node bifurcation, manifesting a phenomenon known as ``critical slowing-down'' \cite{Manneville04}.  Similar energy plateau periods have been reported for PCF \cite{Olvera17}.

Since $E_{k1}\sim 10^{-1}\ll 1$ and $|d^3\sigma/dE_k^3|\ll |d^2\sigma/dE_k^2|$ at $E_k=E_{k1}$ due to the smooth feature of $\sigma$ curves shown in Fig. \ref{fig4}(b), $|E_k-E_{k1}|$ needs not to be much smaller than $E_{k1}$ in order to ignore the higher order terms in Eq. (\ref{eq55}). It is shown in Fig. \ref{fig4}(b) that all the decay rate differences $-(\sigma-\sigma_1)$ at different Reynolds numbers with different initial perturbations ($Re_i$) collapse together and agree well with the quadratic relation Eq. (\ref{eq55}) in the range of $0.5E_{k1}<E_k<2E_{k1}$, which covers nearly the whole lifetime of LWP as shown in Fig. \ref{fig4}(c). Consequently,  the lifetime for a case with the initial disturbance kinetic energy $E_k^i$ can be evaluated with Eq. (\ref{eq55}) as
\begin{equation}\label{eq56}
\begin{split}
\tau = \int^0_{E_k^i} \frac {d\ t}{dE_k} dE_k =  \int^0_{E_k^i} \frac {1}{\sigma(E_k)} dE_k \simeq \frac{1}{\sqrt{-\sigma_1 A}}\left[{\rm arctan}\left(\frac{\sqrt{A}(E_k^i-E_{k1})}{\sqrt{-\sigma_1}}\right)+\frac{\pi}{2} \right].
\end{split}
\end{equation}

Note that $E_{k1}/\sqrt{-\sigma_1} \simeq E_{k1}Re_c/\sqrt{(Re_c-Re)\mathcal{D}(E_{k1})}\gg 1$ as $Re$ approaches  $Re_c$. Specifically,
\begin{equation}\label{eq57}
\begin{cases}
 \tau \simeq \frac{\pi}{\sqrt{-\sigma_1A}} \propto (Re_c-Re)^{-1/2} \ &\text{ as } E_k^i-E_{k1}\gg \sqrt{ \mathcal{D}(E_{k1})\delta/(ARe)},\\
\tau_1 =\tau \simeq \frac{\pi}{2\sqrt{-\sigma_1A}}\propto (Re_c-Re)^{-1/2} \ &\text{ as } E_k^i = E_{k1}.
\end{cases}
\end{equation}
Considering that $Re\sim 10^3$ and $\delta \ll 1$, the condition  $E_k^i-E_{k1}\gg \sqrt{ \mathcal{D}(E_{k1})\delta/(ARe)}$ is easy to be satisfied for $E_k^i \ne E_{k1}$ when $\delta$ is small enough. According to Eq. (\ref{eq57}), the lifetime of LWP with high $E_k^i$ should be about twice as long as that with $E_k^i=E_{k1}$, i.e. $\tau/\tau_1 \simeq 2$, a prediction consistent with the simulations shown in Fig. \ref{fig4}(d), indicating that the initial disturbance kinetic energy does affect the lifetime of LWP. 

\section{Growth of LWP} \label{sec: growth}

\begin{figure*}
	\centerline{\includegraphics[scale=0.125]{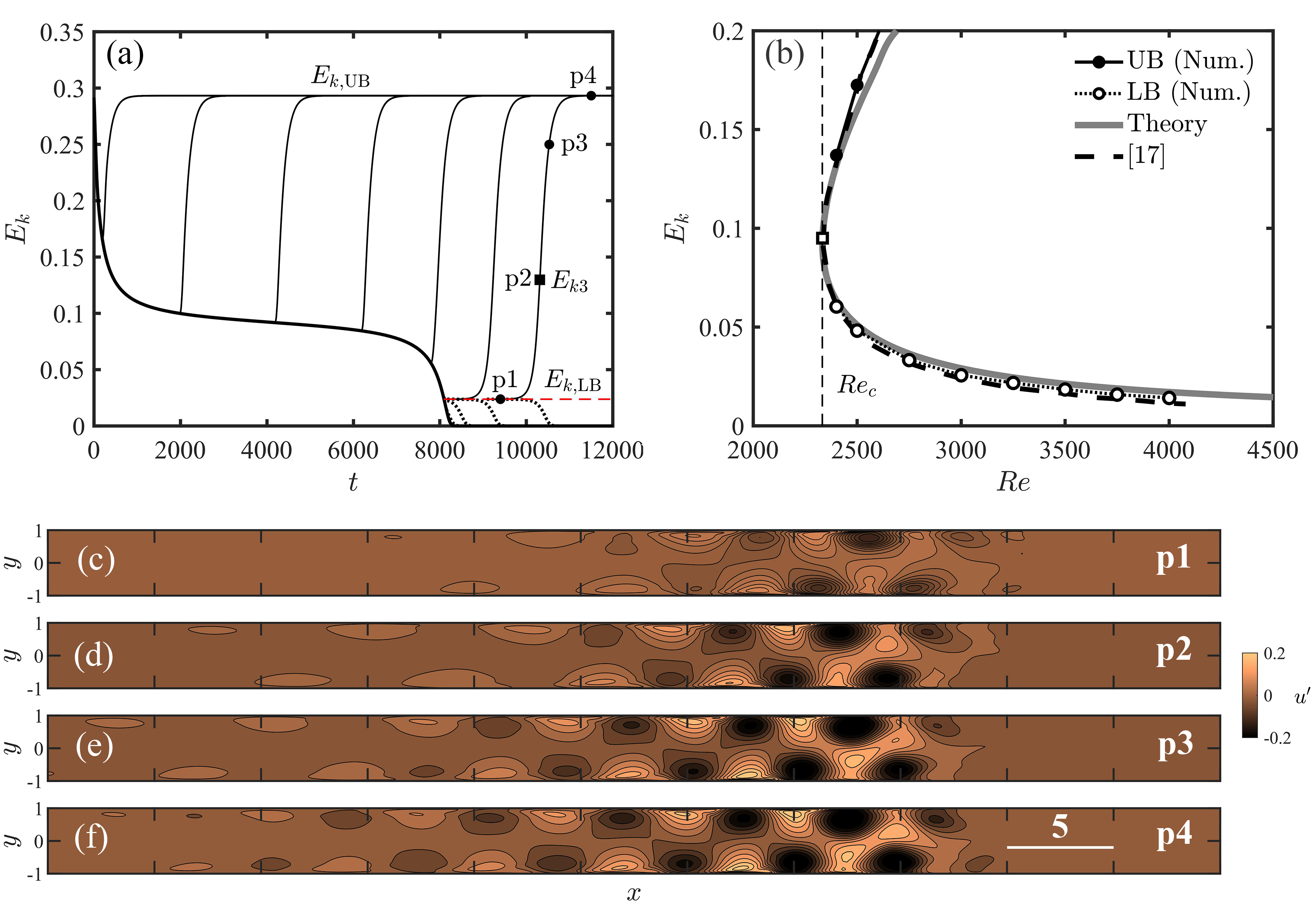}}
	\caption{(a) $E_k$ of growing and decaying LWPs at $Re=3000$ shown by solid and dotted lines, respectively. The thick solid curve represents the $E_k$ of the decaying case shown in Fig. \ref{fig3}(a-d),  and $E_{k3}$ indicates the inflection point. UB and LB represent the upper branch and the lower branch, respectively. (b) Numerically obtained upper branch (filled circle) and lower branch (empty circle) in comparison with the theoretical predictions (grey curve), which corresponds to the $\mathcal{D}/\mathcal{P}$ curve shown in Fig. \ref{fig3}(g). The empty square indicates the turning point with $Re=Re_c$.  (c-f) Iso-contours of $u'$ at different times shown in (a).
 }
	\label{fig5}
\end{figure*}

\begin{figure*}
	\centerline{\includegraphics[scale=0.125]{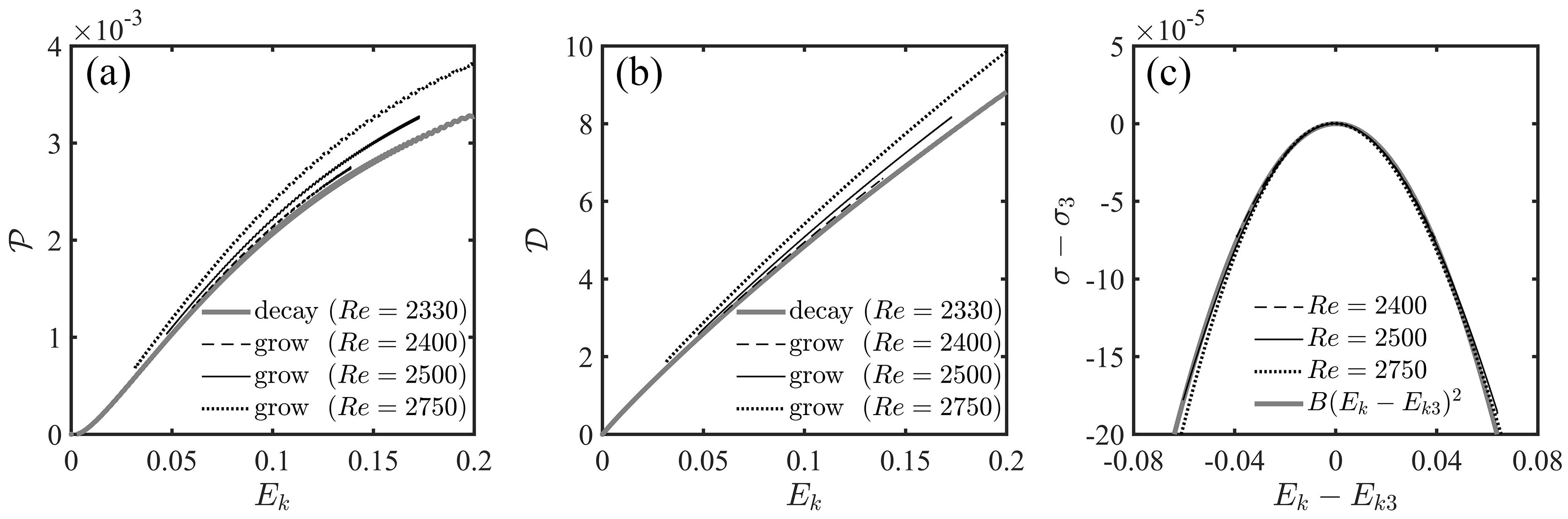}}
	\caption{(a) The energy production term $\mathcal{P}$ and (b) the disturbance enstrophy $\mathcal{D}$ of LWPs growing from the lower branch to the upper branch at  different $Re$s. The corresponding data for a decaying LWP, whose initial field is obtained at $Re_i=3000$,  are drawn as well for reference. (c) Growth rate difference $\sigma-\sigma_3$ as a function of $E_k-E_{k3}$. The fitting coefficient $B=-0.0496$.
}
	\label{fig6}
\end{figure*}

When $Re>Re_c$, it is shown in Fig. \ref{fig5}(a) that the weak LWPs, which are obtained numerically at different times during a decay process, grow directly to a saturated state with the same $E_{k,\mathrm{UB}}$, the corresponding upper-branch LWP solution. Especially, when $E_k$ of the initial LWP is lower than a specific value $E_{k,\mathrm{LB}}$, the LWP will decay as shown by dotted curves in Fig. \ref{fig5}(a), where the bisection method \cite{Skufca06} is used  to track the lower-branch or edge state. The lower and upper branches of LWP calculated by the time-dependent simulations are shown by the empty and filled circles in Fig. \ref{fig5}(b), respectively, where the solutions obtained by the continuation method \cite{Zammert17} are shown as references. 

Both branches of LWP are steady solutions, and according to Eq. (\ref{eq51}) the corresponding $\sigma=0$ and $\mathcal{D}/\mathcal{P}=Re$. The horizontal dashed line in Fig. \ref{fig3}(g) is tangent to the $
\mathcal{D}/\mathcal{P}$ curve, indicating the minimum $Re$ for steady LWP. Moving this line up will intersect with the $\mathcal{D}/\mathcal{P}$ curve at two points, whose $E_k$s should correspond to $E_{k,\mathrm{LB}}$ and $E_{k,\mathrm{UB}}$, respectively. Based on the $\mathcal{D}$ and $\mathcal{P}$ obtained during the decay process ($Re<Re_c$), the predicted $E_k$s for the upper and the lower branches are shown by the grey curve in Fig. 5(b), agreeing with the simulation data. Especially, the  predicted $E_{k,\mathrm{LB}}$ is consistent with the simulation value not only near $Re_c$ but also at high Reynolds numbers, e.g. $Re=4000$, indicating that the pattern preservation approximation is applicable for LWPs with low $E_k$s in a wide range of $Re$. The flow patterns at different times during a growth process from the lower branch to the upper branch are shown in Figs. \ref{fig5}(c-f) and share the same features, localized short-wavelength waves with a strong downstream end and a long and decaying upstream tail.
 
The kinetic energy production term $\mathcal{P}$ and the disturbance enstrophy $\mathcal{D}$ during the growth processes at different Reynolds numbers are shown in Figs. \ref{fig6}(a) and \ref{fig6}(b), respectively. The curves for moderate Reynolds numbers agree acceptably with each other, confirming the availability of the pattern preservation approximation. When both the Reynolds number and the disturbance kinetic energy are high, e.g. $E_k > 0.15$ and $Re > 2500$, discrepancies among the $\mathcal{P}$ and $\mathcal{D}$  curves of growing cases can be observed, reflecting the Reynolds number effect on strong LWPs and the limitation of the approximation. When the decay process starts from a LWP saturated at a higher $Re$, there is always an offset adaptation. This offset effect becomes significant as $E_k>0.15$, and leads to the underestimation of $\mathcal{P}$ and $\mathcal{D}$ [Fig.  \ref{fig6}(a) and \ref{fig6}(b)], the dependence of  decay rate on the initial $E_k(Re_i)$  [Fig. \ref{fig4}(b)],   and the underprediction of the upper branch with high $E_k$ as shown in Fig. \ref{fig5}(b).   

When LWP grows from the lower branch to the upper branch, there is one and only one inflection point in the $E_k$ curve, corresponding to a disturbance kinetic energy $E_{k3}$ with the largest growth rate $\sigma_3$ as shown in Fig. \ref{fig5}(a). This phenomenon can be explained briefly with the pattern preservation approximation as follows. At the inflection point, Eq. (\ref{eq53}) is still applicable, i.e. $d\mathcal{D}/d\mathcal{P}=Re$, and according to Fig. \ref{fig3}(g) there is only one intersection point between a horizontal line for $Re>Re_c$ and the $d\mathcal{D}/d\mathcal{P}$ curve in the region above the $\mathcal{D}/\mathcal{P}$ curve, which confines the $E_k$ range between the lower and the upper branches. Using the similar arguments to those for Eq. (\ref{eq55}), the growth rate can be expanded around this inflection point as well and we have $\sigma \simeq \sigma_3 +B(E_k-E_{k3})^2$, where $B$ may be further expanded as $B= (1/2) (d^2 \sigma/d E_k^2)_{E_k=E_{k3}}= (1/2) (d^2 \sigma/d E_k^2)_{E_k=E_{kc}}+...$. As shown in Fig. \ref{fig6}(c), the growth rate differences $\sigma-\sigma_3$ obtained numerically at different Reynolds numbers coincide with each other and agree with the quadratic relation. The fitted coefficient $B=-0.0496$, which is very close to -0.048, the value of  $(1/2) (d^2 \sigma/d E_k^2)_{E_k=E_{kc}}$. 

According to the above discussions, just based on the pattern preservation approximation and the $\mathcal{D}$ and $\mathcal{P}$ data obtained from the decaying cases, both $Re$ and $E_k$ of the whole lower branch, the turning point, and the upper branch with $E_k<0.15$ of this saddle-node bifurcation can be predicted directly by the disturbance kinetic energy equation. The reason why the flow pattern can be preserved during the decay and growth processes lies in the nonlinear self-sustaining mechanism of LWP \cite{Xiao2021self}.  In the mean-flow field after deducting the basic flow, there is a finite-amplitude vortex dipole at the downstream end of LWP, which provides an unstable region to produce TS-type waves propagating and decaying in the upstream stable region, forming the long upstream tail. On the other hand, the Reynolds stresses contributed by the TS-type waves strengthen the vortex dipole and hence LWP can be self-sustained. Since the length of vortex dipole is confined by the channel height and the wavelength of the TS-type wave in the long upstream tail are mainly determined by the dispersion relation of the basic flow, i.e. both length scales are intrinsically related to the channel height, the geometric shape of LWP weakly depends on the Reynolds number, and its pattern can be retained.

\section{Conclusions}\label{sec: conclusions}
Near the onset of turbulence in three-dimensional PCF and PPF, the growth of disturbance kinetic energy may be caused by band extension and band split, and the decrease of $E_k$ may correspond to band breaking or longitudinal contraction. For two-dimensional PPF, however, the variation of $E_k$ does not change much the flow pattern of LWP. By applying the pattern preservation approximation, where the integral kinematic properties are  independent of $Re$ and only functions of $E_k$, it is shown that the disturbance kinetic energy equation can be transformed into the classical differential equation for saddle-node bifurcation. As a result, LWP's lifetime can be derived analytically and is shown to be proportional to $(Re_c-Re)^{-1/2}$, a scaling law found in numerical simulations. Especially, the Reynolds number and  $E_k$ of both the upper branch ($E_k<0.15$) and the lower branch  predicted by the disturbance kinetic energy equation are consistent with the simulation data in a wide range of Reynolds number, indicating that the pattern preservation is an intrinsic character of LWP. Considering that the two-dimensional PPF is a simplified model for the three-dimensional case and retains required ingredients for the subcritical transition, e.g. solitary localized structure at moderate Reynolds numbers, structure split at relative high Reynolds numbers, and small-scale quasi-periodic components coupled with a large-scale mean flow, the present results are expected to be helpful in understanding the fully three-dimensional transitions in channel flows and wait for experimental validations through film or other two-dimensional flow techniques.

\begin{acknowledgements}
 This work benefits from the insightful discussions with P. Manneville, J. Jim\'{e}nez, G.W. He, and Y. Wang.  The code SIMSON from KTH and help from P. Schlatter, L. Brandtl and D. Henningson are gratefully acknowledged. The simulations were performed on TianHe-1(A), and the support from the National Natural Science Foundation of China is acknowledged (Grants No. 91752203 and 11490553).
\end{acknowledgements}

\bibliography{pp}% Produces the bibliography via BibTeX.

\end{document}